\def   \ni {\noindent}
\def   \bsk {\vskip 15truept}

\documentstyle[epsfig]{article}
\begin{document}

\font\affiliation=cmssi10
\font\author=cmss10
\font\caption=cmr8
\font\references=cmr8
\font\title=cmssbx10 scaled\magstep2
\def\ref{\par\noindent\hangindent 15pt}
\null

\title{\ni Probing the Central Engine 
             of the Narrow-Line Seyfert 1 Galaxies
 }
                                
\bsk \bsk
\author{\ni A.~Janiuk $^{1}$, J.~Kuraszkiewicz $^{2}$, B.~Czerny $^{1}$ }

\bsk
\affiliation{1) N. Copernicus Astronomical Centre,  Bartycka 18, 00-716, Warsaw, Poland\\

2) Harvard Smithsonian Center for Astrophysics, 60 Garden Street, Cambridge, MA 02138, USA
}       
\bsk
\baselineskip = 12pt

\abstract{\ni

The central engine of the the Narrow Line Seyfert 1 galaxies
is being probed. We use the ASCA and RXTE data to model the X-ray 
primary continuum as well as the reflected component and iron $K\alpha$ line.
Since these are strongly coupled, we obtain independent measurements of the 
disc ionization level and the orientation dependent reflection amplitude.
Using the available Optical/UV data we also estimate the black hole masses 
and the  $L/L_{Edd}$ ratios,
 which are probably related to 
Boroson \& Green eigenvector~1.

}                                                    
\bsk
              
\section{Introduction}

Studies of optical emission lines in quasars have revealed 
strong correlations between emission line properties, that are
possibly related to the central accreting black hole system. Boroson
\& Green (1992) used the Bright Quasar Sample (BQS) and identified a
set of optical emission line properties that vary together (optical
FeII and [OIII]\,$\lambda 5007$ strengths, H$\beta$ width and blue
asymmetry), called the Boroson \& Green eigenvector~1 (EV1). Eigenvector 1
 was found by Boller \& Brandt (1998) to correlate with X-ray properties
($\alpha_x$, L$_{2keV}$).  As the X-rays originate in the vicinity of
the central black hole (at distances $< 100 R_{g}$), hence eigenvector 1
 is possibly linked to and driven by the central engine. 
To find the parameters of the central engine that drive eigenvector 1,
 we examined first objects with extreme EV1 properties.  A class of
Narrow-Line Seyfert~1 galaxies (NLS1s) is found to lie at the low EV1
end of the Bright Quasar Sample studied by Boroson \& Green. 
These objects exhibit particularly narrow H$\beta$ line, of FWHM$ < $2000 km/s,
strong Fe\,II emission and [O\,III]/H$\beta$ $<$3.
Narrow Line Seyfert 1 galaxies, when compared
to typical AGN, show hotter and more pronounced big blue bumps and
steeper soft-X-ray slopes. This can be explained due to higher
ratios of their luminosity to the Eddington luminosity 
(e.g. Pounds et al. 1995), however there have also been attempts to explain 
the extreme properties and variability of the NLS1 class by their high 
inclination angles to the line of sight (e.g. Brandt \& Gallagher, 2000).   
 This points towards either $L/L_{Edd}$ or orientation as the
primary driver of eigenvector 1.

For large accretion rates the matter of the accretion disk is more ionized,
 leading
to an increase of the K$\alpha$ line centroid energy above the neutral case
value of 6.4 keV (see Matt, Fabian \& Ross 1993, \.{Z}ycki \& Czerny
1994). This fluorescent line is linked to the Compton reflection feature, 
as they are both resulting from the irradiation of the disk surface by the 
hard X-rays. Therefore, in the spectral analysis we used the model of 
the reflection hump plus the iron line (\.Zycki, Done \& Smith, 1997).

\bsk
              
\section{Results of the spectral analysis}

Both literature and archives (NED, HEASARC) were searched to obtain a
complete UV to hard-X-ray continuum for each source. The bolometric
luminosity $L_{bol}$ was estimated using the extrapolation of V
magnitude to the 2500\AA\  bandpass, with the assumption of the continuum
slope of 0.5, and taking into account the X-ray luminosity 
$L_{2-10keV}$ derived from the spectral fits.  Then we calculated the
accretion rate assuming the accretion efficiency of 1/16 and the
central black hole mass was estimated from the standard relation
$F_{\nu} \sim \nu^{1/3}(M\dot M)^{2/3}$. To calculate distances we assumed H=50 km s$^{-1}$ Mpc$^{-1}$ and q=0.

%--------------------------  table 1

\begin{table}
\begin{center}
\caption{Luminosities and black hole masses}\vspace{1em}
%\tablewidth{100 mm}
    \renewcommand{\arraystretch}{1.2}
    \begin{tabular}[h]{lccccr}
      \hline
Object name & $log \nu L_{\nu}(2500 \AA)$ & $log L_{2-10 keV}$ &
$L/L_{Edd}$ & $\dot M$  & $log M_{BH}$ \\
            & [erg/s] & [erg/s] &  & [$M_{\odot}/yr$] & [$M_{\odot}$] \\
      \hline
PKS 0558-504     & 45.31 & 44.66 & 0.33 & 3.345 & 8.45 \\
IRAS 13349+243   & 45.11 & 44.17 & 0.20 & 1.84  & 8.40 \\
PG 1211+143      & 45.01 & 44.59 & 0.27 & 1.79  & 8.27 \\
ARK 564          & 43.97 & 43.71 & 0.10 & 0.18  & 7.70 \\
      \hline \\
      \end{tabular}
    \label{tab:table1}
  \end{center}
\end{table}

%\tablenotetext{a}{in erg/s}
%\tablenotetext{b}{in $M_{\odot}/yr$}
%\tablenotetext{c}{in solar masses}
%\tablecomments{To calculate distances we assumed H=50 km s$^{-1}$ Mpc$^{-1}$ and q=0.}
%\enddata

%\end{deluxetable}

%---------------------------------

The value of eigenvector 1 has been determined by Boroson \& Green for only  
one quasar PG1211+143. For the other 
three objects we estimated this value by comparing the EV1 range
of objects with similar values of FWHM~H$\beta$, Fe\,II/H$\beta$
and [O\,III] strength in the Boroson \& Green 
PG sample. In Table 2  we show the minimum and maximum values of eigenvector 1
 corresponding to the quasars with similar properties to our objects 
(up to 10\% difference in FWHM~H$\beta$), and the mean value $<EV1>$, weighted 
by the two other crucial properties: Fe\,II/H$\beta$ and [O\,III]. The values 
of H$\beta$ width, Fe\,II/H$\beta$ and 
[O\,III]/H$\beta$ were taken from Boller et al. 1996 and Leighly 1999

%--------------------------  table 2
\begin{table}
\begin{center}
\caption{Emission line properties}
    \renewcommand{\arraystretch}{1.2}
    \begin{tabular}[h]{lccccccr}
      \hline
Object name & FWHM H$\beta$ & Fe\,II/H$\beta$ & [O\,III]/H$\beta$ &
$EV1_{min}$ & $EV1_{max}$   & $<EV1>$ \\

      \hline
PKS 0558-504     & 1500 & 1.56 & 0.04 & -5.92 & -2.93 & -4.77  \\
IRAS 13349+243   & 2100 & 6.5  & 0.13 & -5.39 & 2.07  & -4.67 \\
PG 1211+143      & 1900 & 0.52 & 0.14 &   -   & -     &  0.464\\
ARK 564          & 720  & 0.8  & 0.96 & -6.36 & -2.41 & -2.41 \\
%\tablecomments{The values of H$\beta$ width, Fe\,II/H$\beta$ and 
%[O\,III]/H$\beta$ were taken from Boller et al. 1996 and Leighly 1999}
      \hline \\
      \end{tabular}
    \label{tab:table2}
  \end{center}
\end{table}

%---------------------------------

The X-ray properties were examined by fitting simultaneously the ASCA and RXTE
 data, using the XSPEC ver. 10.0. We used the power law continuum model, 
corrected for the galactic absorption. Then we added the reflection component,
 which in most cases gave a significant improvement in the fit. This spectral 
feature is parameterized by the ionization parameter $\xi$, reflection 
amplitude $\Omega/2\pi$, inner disc radius $R_{in}$ and the disc 
inclination $\cos i$. The incident hard X-ray flux was assumed to have a 
radial distribution fixed at $F_{irr} \sim r^{-3}$.

In Table 3 we show the results of spectral fitting to the X-ray data.
The model ingredients for PKS 0558-504, IRAS 13349+243 and  ARK 564 are: 
galactic absorption, power law and reflected component with iron line.
The model ingredients for PG 1211+143 are: galactic absorption, comptonized 
black body, power law and reflected component with iron line

%--------------------------  table 3
\begin{table}
\begin{center}
\caption{Results of simultaneous fitting to ASCA and RXTE data}
    \renewcommand{\arraystretch}{1.2}
    \begin{tabular}[h]{lcccccr}
      \hline

Object name & $\Gamma$ & $\Omega/2\pi$ & $\xi$ & $\cos i$ & $R_{in}$ [$R_{g}$]&
 $\chi^{2}_{\nu}(d.o.f.)$ \\

      \hline

PKS 0558-504  & $2.4 \pm 0.02$ & $1.5 ^{+0.5}_{-0.8}$ &
 $10^{+130}_{-10}$ & $0.3^{+0.2}_{-0.1}$ & $6.0^{f}$ & 1.0 (714) \\
IRAS 13349+243  & $2.21 \pm 0.09$ & $ 0.3^{+1.6}_{-0.25}$ &
 $160^{+1500}_{-60}$ & $0.5^{f}$ & $6.0^{f}$ & 0.96 (127) \\
PG 1211+143  & $2.18 \pm 0.03$ & $0.85 ^{+0.65}_{-0.55}$ &
 $500^{+600}_{-450}$ & $0.9 \pm 0.1$ & $6.0^{+4.0}_{-0}$ & 1.04 (535) \\
ARK 564  & $2.68 \pm 0.05$ & $0.96 \pm 0.36$ &
 $525^{+1400}_{-370}$ & $0.8^{f}$ & $6.0^{+7.0}_{-0}$ & 0.91 (1029) \\

%\tablenotetext{a}{Inner radius in $R_{g}$}
%\tablenotetext{b}{The model ingredients are: galactic absorption, power law and reflected component with iron line}
%\tablenotetext{c}{The model ingredients are: galactic absorption, comptonized black body, power law and reflected component with iron line}
%\tablecomments{The model consists of  the Comptonized black body,
%power law, Galactic
%absorption (fixed value) and reflection component with an iron line.}
      \hline \\

      \end{tabular}
    \label{tab:table2}
  \end{center}
\end{table}

%---------------------------------

\bsk
              
\section{Conclusions}

Preliminary results showed that Narrow Line Seyfert 1 galaxies have high 
$L/L_{Edd}$ ratios
and relatively small black hole masses when compared to normal Seyfert
1 galaxies.
%*****podaj tu liczby jakie sa wartosci dla NLS1 a jakie dla Sy1 
The mass determination by means of the power density spectra 
(Czerny et al. 2001) indicates, that in case of Seyfert 1 galaxies 
$log M_{BH}$ is $\sim 7.5$, while luminosity is only about 2-5\% of the 
Eddington luminosity.  For NLS1 the corresponding values are $log M_{BH} \sim 6.5 - 8.2$ and $L\sim 20-40\% L_{Edd}$. 
Since NLS1 have small eigenvector 1 values this finding points to $L/L_{Edd}$
ratio (or  accretion rate) as the primary driver of
eigenvector~1 (see also Boroson 2001). This also points to $L/L_{Edd}$ as the main physical
parameter responsible for the extreme properties of the Narrow Line
Seyfert 1 galaxies.

For larger $L/L_{Edd}$ the matter of the accretion disk is more
ionized, having an impact on the reflected spectrum shape. In our
sample for only PKS 0558-504 the neutral reflection was acceptable at
90\% confidence level, while for the other three objects the fit required
the disk surface to be mildly ionized.  The disk seems to extend down
to the marginal stable orbit ($R_{in}\sim 6.0 R_{g}$) in all
objects. However, the results for the inclination angle do not seem to favor
any particular orientation.

\bsk
\baselineskip = 12pt

{\references \ni REFERENCES
\bsk

\ref Boller T., \& Brandt, W.N., 1998, Astron. Nachr., 319, 7

\ref Boller T., Brandt, W.N, Fink H., 1996, A\&A, 305, 53 

\ref Boroson, T.A., \& Green, R.F., 1992, ApJS, 80, 109

\ref Boroson, T.A., 2001, astro-ph/0109317

\ref Brandt W.N., Gallagher S.C., 2000, NewA Rev., 44, 461

\ref Czerny B., Nikolajuk M., Piasecki M., Kuraszkiewicz J., 2001, MNRAS, 325, 865

\ref Janiuk A., \. Zycki P.T., Czerny B., 2000, NewA Rev., 44, 1003

\ref Kuraszkiewicz J., Wilkes B., Czerny B., Mathur S., 2000, ApJ, 542, 692

\ref Leighly K., 1999, ApJS, 125, 317 

\ref Matt, G., Fabian A.C., Ross, R.R., 1993, MNRAS, 262, 179

\ref Pounds K., Done C., Osbourne J., 1995, MNRAS, 277, L5

\ref \.{Z}ycki P. \& Czerny B., 1994, MNRAS, 266, 653

\ref \. Zycki P.T., Done C., Smith D.A., 1997, ApJ, 488, L113
}

\end{document}